\documentclass[sigconf]{acmart}

\AtBeginDocument{%
  \providecommand\BibTeX{{%
    \normalfont B\kern-0.5em{\scshape i\kern-0.25em b}\kern-0.8em\TeX}}}

\copyrightyear{2024}
\acmYear{2024}
\setcopyright{rightsretained}
\acmConference[DaMoN '24]{20th International Workshop on Data Management
on New Hardware }{June 10, 2024}{Santiago, AA, Chile}
\acmBooktitle{20th International Workshop on Data Management on New
Hardware (DaMoN '24), June 10, 2024, Santiago, AA, Chile}
\acmDOI{10.1145/3662010.3663447}
\acmISBN{979-8-4007-0667-7/24/06}

\makeatletter
\gdef\@copyrightpermission{
  \begin{minipage}{0.3\columnwidth}
   \href{https://creativecommons.org/licenses/by/4.0/}{\includegraphics[width=0.90\textwidth]{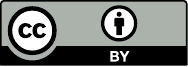}}
  \end{minipage}\hfill
  \begin{minipage}{0.7\columnwidth}
   \href{https://creativecommons.org/licenses/by/4.0/}{This work is licensed under a Creative Commons Attribution International 4.0 License.}
  \end{minipage}
  \vspace{5pt}
}
\makeatother

\begin{document}

\title{DuckDB-SGX2: The Good, The Bad and The Ugly within Confidential Analytical Query Processing}

\author{Ilaria Battiston}
\affiliation{
  \institution{Centrum Wiskunde \& Informatica}
  \city{Amsterdam}
  \country{The Netherlands}
}

\author{Lotte Felius}
\affiliation{
  \institution{Centrum Wiskunde \& Informatica}
  \city{Amsterdam}
  \country{The Netherlands}
}

\author{Sam Ansmink}
\affiliation{
  \institution{DuckDB Labs}
  \city{Amsterdam}
  \country{The Netherlands}
}

\author{Laurens Kuiper}
\affiliation{
  \institution{Centrum Wiskunde \& Informatica}
  \city{Amsterdam}
  \country{The Netherlands}
}

\author{Peter Boncz}
\affiliation{
  \institution{Centrum Wiskunde \& Informatica}
  \city{Amsterdam}
  \country{The Netherlands}
}

\renewcommand{\shortauthors}{Ilaria Battiston, Lotte Felius, Sam Ansmink, Laurens Kuiper, \& Peter Boncz}


\begin{abstract}
We provide an evaluation of an analytical workload in a confidential computing environment, combining DuckDB with two technologies: modular columnar encryption in Parquet files (data at rest) and the newest version of the Intel SGX Trusted Execution Environment (TEE), providing a hardware enclave where data in flight can be (more) securely decrypted and processed. One finding is that the "performance tax" for such confidential analytical processing is acceptable compared to not using these technologies. We eventually manage to run TPC-H SF30 with under 2x overhead compared to non-encrypted, non-enclave execution;  we show that, specifically, columnar compression and encryption are a good combination. Our second finding consists of \textit{dos} and \textit{don'ts} to tune DuckDB to work effectively in this environment. There are various performance hazards: potentially 5x higher cache miss costs inside the enclave, NUMA penalties, and highly elevated cost of swapping pages in and out of the enclave -- which is also triggered indirectly by using a non-SGX-aware {\small\tt {malloc}} library. 
\end{abstract}

\maketitle

\section{Introduction and Background}
Confidential Computing aspires to protect data in use by performing computations in a hardware-based, attested Trusted Execution Environment (TEE)~\cite{ccc}. Such TEEs are particularly useful in scenarios where sensitive data is outsourced to an untrusted external cloud provider. The first general-purpose TEE technology introduced in 2015 for mainstream hardware is Intel SGX -- processing queries on e.\:g. AES-encrypted data requires decryption, which exposes sensitive data to malicious actors if they have access to the RAM. 
SGX aims to protect against this threat by separating a dedicated part of the RAM into a secure enclave. 
The enclave keeps data encrypted at all times and only decrypts it transparently inside the CPU package (i.\:e. in the registers and CPU caches), making it inaccessible for untrusted processes.

Intel SGX allows trusted and untrusted software to communicate through ECALLs (enclave calls) and OCALLs (out calls) -- these are, respectively, special invocations from the application to the enclave, and from the enclave to the application. Untrusted software initiates ECALLs to transfer control flow to code inside the enclave. 
Similarly, trusted software uses OCALLs to transfer control flow back to code outside the enclave. 
These interactions involve a CPU context switch and additional steps, such as flushing CPU caches and the Translation Lookaside Buffer (TLB), to maintain the confidentiality of enclave data. 
Communication between trusted and untrusted software is thus expensive and should be avoided when possible.
 
SGX allocates a dedicated memory region, the Processor Reserved Memory (PRM), which is protected from non-enclave accesses.
The PRM contains the Enclave Page Cache (EPC), which in turn stores code and data in encrypted memory pages of 4KB.
To provide an EPC size up to 512GB per socket, the newer version of SGX (SGX2) makes use of the Total Memory Encryption -- Multi-Key (TME-MK), replacing the more expensive Memory Encryption Engine (MEE) used in an older version of SGX.
The TME-MK relies on \texttt{AES-XTS} for encryption, and enables the creation of private memory regions to assure confidentiality.
SGX allows swapping pages in/out of unprotected memory through EPC paging, which occurs when the allocated memory for the PRM is exceeded.
However, this operation is much more expensive than regular page swapping because of additional checks that guarantee the confidentiality and integrity of evicted EPC pages.

This paper does not focus on the security aspect of query processing with SGX -- we are aware of many of its security vulnerabilities\footnote{e.\:g.; https://sgx.fail lists vulnerabilities; www.intel.com/content/www/us/en/security-center/advisory/intel-sa-00329.html is an example mitigation notice.}. 
We think SGX should not be considered a single measure to achieve security but rather a hardening technology, given that all advised mitigations are implemented. Properly operating SGX hardware requires promptly implementing all advised advisories, requiring -- among others -- regular microcode updates, and disabling simultaneous multi-threading.
We hope to transfer some lessons from this work to other/future enclave technologies, including those working on the VM level (Intel TDX~\cite{tdx}, AMDSEV~\cite{sev}, AWS Nitro~\cite{nitro}, ARM CCA~\cite{280904}). Furthermore, we acknowledge additional potential security threats of our implementation and plan to resolve them in the future.

\vspace{2mm}\noindent{\bf SGX databases.}
Multiple database prototypes for Intel SGX1 exist, e.\:g., StealthDB~\cite{DBLP:journals/corr/abs-1711-02279}, EncDBDB~\cite{DBLP:journals/corr/abs-2002-05097}, CryptSQLite~\cite{8247155}, EnclaveDB~\cite{8418608}, and even DuckDB~\cite{16-sam-thesis}. In particular, our previous work describes and implements different approaches for a secure OLAP database, deploying vectorized, compressed, and encrypted execution and placing different components of the engine inside the enclave. 
In all cases, the constrained EPC size of 256 MB significantly affected query performance due to an inefficient page-swapping mechanism.
Consequently, most prototypes for database systems using SGX1 are considered impractical due to severe performance degradation or weak security guarantees.

In SGX on the newer generations of Intel hardware (commonly SGX2), the EPC constraint has been lifted since the maximum size increased to 512GB per socket (with a maximum of two sockets per enclave). This significant advancement in SGX technology opens up new possibilities for secure applications.
Due to its novelty, however, limited research exists on database systems on SGX2. ~\cite{10.14778/3494124.3494146} describes bottlenecks and challenges encountered during the execution of joins in TEEs. ~\cite{10.1145/3533737.3535098} benchmarks SGX2, showing around 25\% overhead compared to an in-memory B-Tree. 
Recent work~\cite{lutsch2024benchmarking} conducts an evaluation of analytical workloads on SGX2, investigating the causes of performance overhead mainly present in hash and radix joins. They argue that a large part of the overhead in SGX2 can be attributed to expensive random memory accesses and a difference in how the CPU executes code inside the enclave. In addition, they warn of possible performance degradation when using internal libraries inside the enclave.

The SGX enclave only allows to secure data while being processed. However, when data exits the enclave, it is again vulnerable to attacks and should be protected. In this paper, we aim to extend the previous research by proposing a fully secure system, efficiently safeguarding not only data in use, but also at rest. 
We run the TPC-H benchmark to perform queries on encrypted Parquet files using DuckDB in SGX2 to provide a secure end-to-end pipeline for analytical queries. We aim to ensure robustness and efficiency, instilling confidence in potential real-world applications.
Therefore, our main research question is: {\em how much does performance degrade inside an enclave, and how to minimize this degradation?}

\section{DuckDB-SGX2}

\begin{figure}[t]
  \centering
  \includegraphics[width=\linewidth]{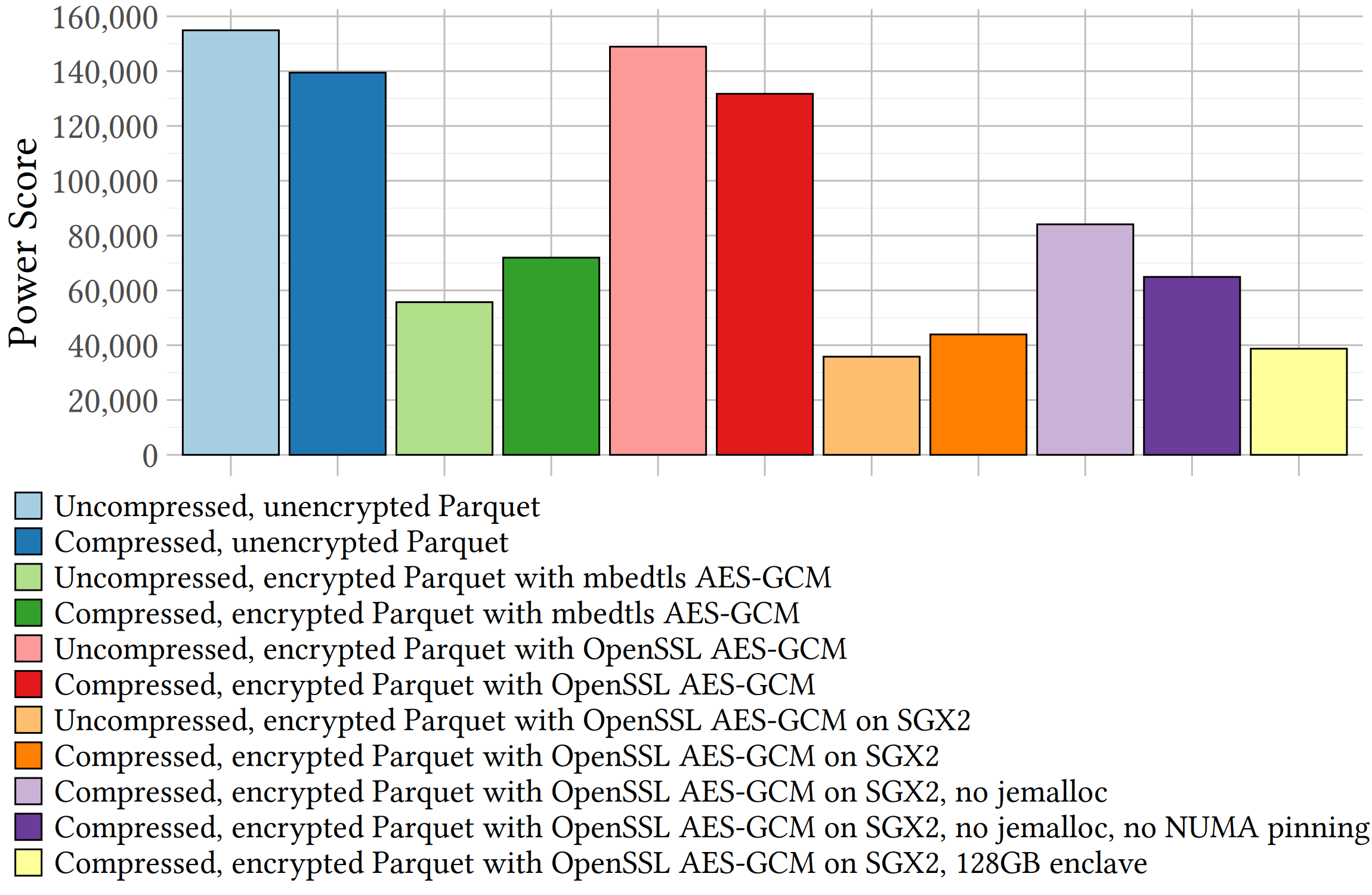}\vspace*{-3mm}
  \caption{DuckDB TPC-H 30GB power scores for various configurations (compression, encryption, SGX); the "good" being the light-purple vs. red (affordable confidentiality); the "bad" light-orange and yellow vs. light-purple (SGX sensitivity to configuration) and the "ugly" choice between blue and light-purple: how much performance is more security worth? }\vspace*{-3mm}
  \label{fig:power}
\end{figure}

Exploiting the larger EPC size, we run DuckDB entirely inside an SGX enclave via Gramine~\cite{10.5555/3154690.3154752}, publishing the manifest in an Open-Source repository\footnote{https://github.com/cwida/DuckDB-SGX2}. 

Gramine (formerly Graphene) is one of the most widely used tools to port applications out-of-the-box to run them inside Intel SGX enclaves. It strives for compatibility with the Linux kernel, providing compatibility with the POSIX standards while remaining minimalistic, implementing only the essential subset of Linux functionality for running portable, hardware-independent applications. It has the capability to run unmodified Linux applications, intercept application requests, and pull the OS functionality inside the enclave. Database systems heavily rely on I/O, scheduling, and memory management to perform their tasks: Gramine supports more than half of the system calls available on Linux, making it feasible to port such workloads\footnote{https://gramine.readthedocs.io/en/stable/} but additionally allows to implement new system calls.

DuckDB~\cite{10.1145/3299869.3320212} is an embedded DBMS for analytical workloads, such as data science and data transformation pipelines. 
It can seamlessly operate within mobile apps, in-browser environments (utilizing WASM), as well as on laptops and in cloud computing settings. This flexibility is made possible by its compact footprint, portable code, and lack of dependencies. Its in-process nature allows for easy porting in Gramine since DuckDB can run directly within the application. 
It is easily possible to build cloud services using DuckDB by embedding it in a server process that accepts queries from the network and sends result sets back.  
Our study also intends to inform about the efficiency of such architectures if they were to use SGX2 for security hardening.

DuckDB implements compressed vectorized execution, allowing direct queries on Parquet files and recently added support for its encryption.
Specifically, a Parquet file consists of different modules: pages, headers, column and offset indexes, and a footer.
Parquet Modular Encryption encrypts each of these separately, using AES-GCM or AES-CTR (although the latter can only be used to encrypt data pages).

For portability reasons, the DuckDB core system, which includes its Parquet functionality, does not make use of assembly or intrinsics, as these are platform-specific; instead, DuckDB inlines the pure C++ {\small\tt mbedtls} library for dealing with encrypted data.
However, ARM and X86 have specialized CPU instructions for AES (not auto-generated by compilers), which are not used for Parquet [de/en]cryption in DuckDB. We, therefore, modified the Parquet functionality in DuckDB to switch at runtime to the {\small\tt OpenSSL} implementation of encryption, exploiting dedicated AES CPU instructions whenever available. We show that this strongly improves encryption and decryption performance.

However, during our preliminary experiments, we struggled to explain a significantly deteriorating performance after running multiple queries requiring large hash tables (e.\:g. Q18). Our investigation eventually led us to disable the {\small\tt jemalloc} memory allocator that DuckDB uses by default on Linux (as recommended in the annals of DaMoN~\cite{10.1145/3329785.3329918}). We attribute the better performance of the {\small\tt glibc} {\small\tt malloc} implementation to its perceived ability to give back unused memory to the OS without incurring EPC paging, i.\:e., by avoiding encryption of freed pages.

\vspace{2mm}\noindent{\bf Evaluation.}
Our platform is a bare-metal instance provided by Intel, running Ubuntu 23.10 with two Intel Xeon Platinum 8570 CPUs (224 cores) and 1TB RAM across 4 NUMA regions.
In order to obtain the best performance, the data should fit in the enclave, the enclave should fit in the EPC, and the EPC should be NUMA-local: hence we run TPC-H SF30 on DuckDB 0.10 with {\small\tt SET memory\_limit='40GB'} and {\small\tt SET threads=56}, requesting a 64GB enclave via Gramine 1.6, with 256GB EPC. The EPC is spread across the NUMA nodes; however, the memory and cores used by Gramine are restricted to one region using {\small\tt numactl}. 

Figure~\ref{fig:power} starts with uncompressed, unencrypted Parquet (light-blue). Adding compression (blue) slightly affects performance, but instead, adding {\small\tt mbedtls} encryption (light-green) causes a 3x slowdown.
Combining both (green) reduces the penalty to 2x, as less data needs to be decrypted thanks to compression. 
Enabling the use of AEX decryption CPU instructions (i.\:e. {\small\tt OpenSSL}) turns the tables, as uncompressed encrypted (pink) then becomes only barely faster than compressed encrypted (red) and has as little as 10\% overhead in the TPC-H power score over uncompressed unencrypted (light-blue). However, running this in SGX constitutes a performance disaster due to the EPC paging caused by {\small\tt jemalloc} memory fragmentation (light-orange).

We attribute this behavior to the use of {\small\tt mmap} calls by {\small\tt jemalloc}: each of these calls requires to zero the corresponding memory region, to provide default values to the memory application. 
Modern operating systems implement optimizations to speed up this operation, such as inserting all-zero pages. 
However, SGX is not able to take advantage of such strategies as it does not have access to the page tables. 
This limitation leads to the usage of {\small\tt memset(0)}, which also incurs additional overhead.

Switching to the {\small\tt glibc malloc} brings us to the best SGX configuration (light-purple), which only pays a 50\% "security tax" for SGX alone (red). 
We attribute the remaining overhead to expensive cache misses and random memory accesses, which occur frequently in e.\:g. hash tables in joins and aggregations, and are known to be less performant in Intel SGX \cite{lutsch2024benchmarking}. 
The performance degradation however is still acceptable; the overhead is less than 2x compared to no security (light-blue). 
Note that this optimization is only feasible when the enclave is larger than {\small\tt 64MB * 56 threads}, as per-thread arenas might consume a large portion of enclave memory. Furthermore, {\small\tt malloc} is vulnerable to memory fragmentation, and it can lead to out-of-memory errors quickly. We believe that {\small\tt mimalloc}\footnote{\href{https://github.com/microsoft/mimalloc}{https://github.com/microsoft/mimalloc}} could be a better solution which suits both SGX and non-SGX query processing, and we plan to evaluate this approach in our future research. The roadmap also includes investigating and implementing improvements in the way DuckDB employs {\small\tt jemalloc} in order to decrease the amount of {\small\tt mmap} and {\small\tt munmap} calls.

Letting go of the NUMA locality (dark-purple) also causes a slowdown, and letting the DuckDB memory overrun the EPC size (yellow) induces EPC paging and performance disasters. Compression reduces the memory footprint (orange) but is still quite inadequate.

\begin{figure}[th]
  \centering
  \includegraphics[width=\linewidth]{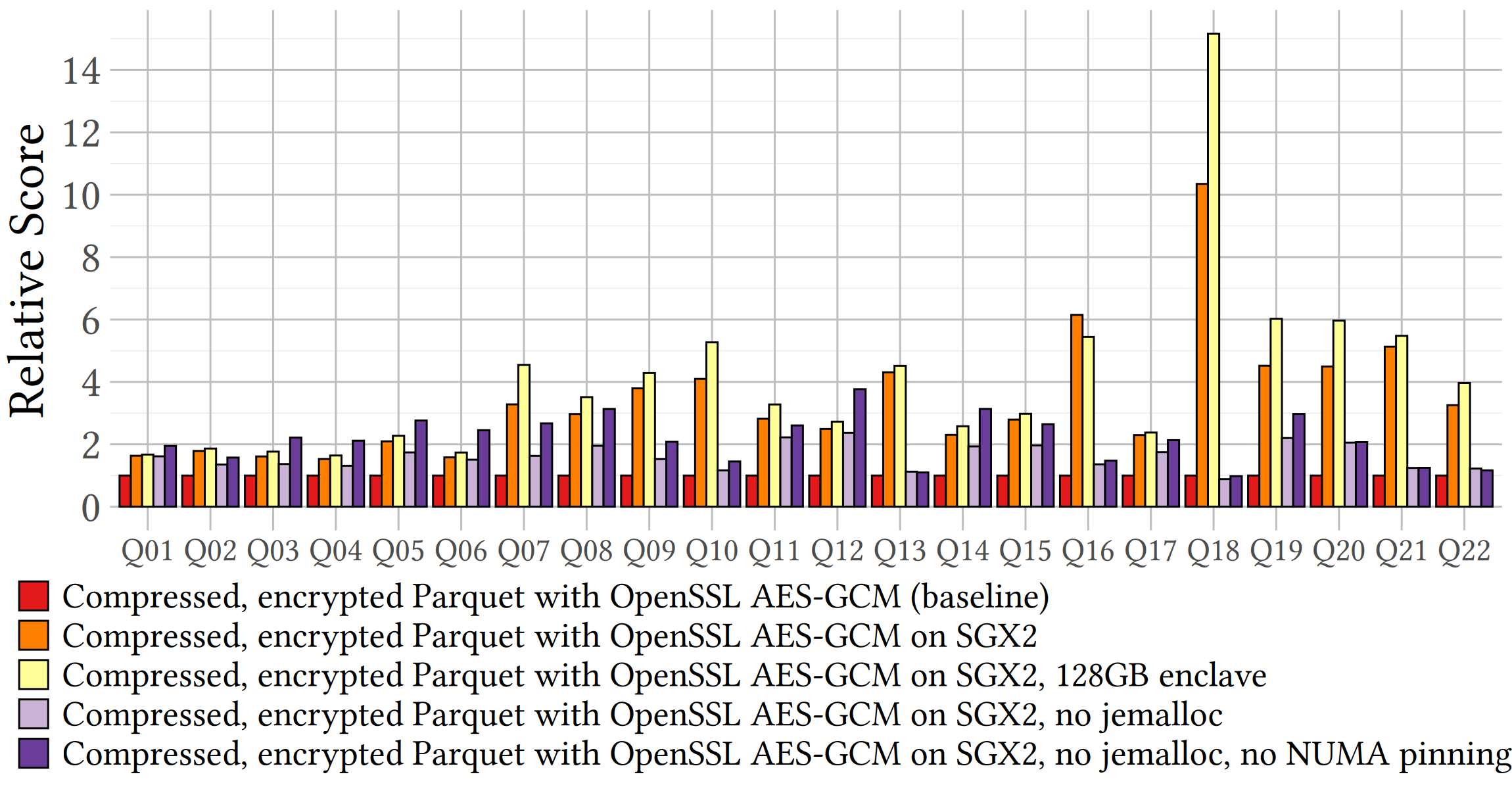}
  \caption{Relative score of TPC-H SF30 (average of 5 runs, compared to the encrypted Parquet {\small\tt mbedtls} baseline --- lower is better). This goes up to 16x (yellow) when configurations are not carefully optimized; however, in the best-case scenario (light purple), each query suffers from at most 2x overhead, a tradeoff we consider acceptable in order to protect our data. Furthermore, the overhead of SGX varies significantly over different queries --- we attribute this to the higher cache miss cost.}
  \label{fig:queries}
\end{figure}

Figure \ref{fig:queries} shows a more detailed overview of the previously mentioned "bad" effects, compared with the "good" performance of the appropriate configuration: in particular, we point out the impact of paging when running different queries consecutively. 

Finally, we look deeper into the performance bottlenecks of single query operators. We denote that there are some queries whose overhead is more significant, as noted in Figure \ref{fig:queries}: the queries with the worst relative score (Q05, Q08, Q11, Q12, Q19) involve either large aggregations or large joins. These operations, in general, are typical for OLAP workloads and revolve around hashing, one of the most expensive processes in terms of cache misses. When performing the TPC-H queries on our machine, we incur up to 70\% misses for each query among all cache references. 

This issue has already been assessed by previous work~\cite{10.14778/3494124.3494146} ~\cite{lutsch2024benchmarking}, denoting the bottleneck of the join performance on SGX. Furthermore, aggregation workloads with many unique groups are suboptimal for hash aggregation on SGX. DuckDB's hash aggregation is inspired by HyPer's parallel hash aggregation~\cite{morsel}. In the first phase of the aggregation, as described in~\cite{pub:34114}, each thread pre-aggregates, i.\:e., duplicates, data in a small, fixed-size hash table that fits in the CPU cache. When this hash table is full, it is reset, and the data is radix-partitioned. This process repeats until all input data has been read. The small fixed-size hash table keeps cache misses to a minimum while still being able to efficiently reduce frequently occurring groups. After the first phase, the radix-partitions are exchanged. In the second phase, each thread performs a partition-wise aggregation. During this phase, however, the hash table can no longer be a small, fixed size; it must grow to fit all unique groups in a partition. If the input contains many unique groups, the hash table will be large, and the probability of cache misses increases, diminishing lookup efficiency. Additionally, cache misses, along with random access, seem to be another source of overhead in SGX, and therefore amplify the performance degradation: future work should deeply investigate this behaviour. 


The number of ECALLS and OCALLS is furthermore in the order of magnitude of hundreds of thousands for each query: since database systems heavily rely on I/O and memory copy, it is essential to investigate more efficient methods such as custom syscalls to minimize this overhead, avoid unnecessary data movement, and [en/de]cryption between disk and memory. In general, data movement between CPU and encrypted memory seems to cause another performance bottleneck due to the write amplification of encrypted pages and their metadata.

DuckDB has a buffer manager to cache blocks of its internal table format but does not use this for Parquet, which is an external format. We note that if data at rest is already encrypted, the buffer manager could make use of memory outside of the enclave to avoid performance overhead incurred by mechanisms inside the enclave ~\cite{16-sam-thesis}.


\section{Security Assessment}\label{sec}
In this section, we briefly discuss the security assessment of our implementation. As we previously mentioned, this paper's primary focus is to provide an overview of the main bottlenecks of analytical workloads on SGX2; however, for real-world usage, it is necessary to understand the security model and its guarantees.

For data at rest, disk-based encryption serves as a foundational layer in mitigating potential security threats, provided that proper storage and management of encryption keys are ensured, a consideration momentarily ignored in our design. While encrypted files protect against unauthorized access, they introduce challenges regarding data modification. Partial modifications to encrypted data necessitate either the re-encryption of the entire database system or the implementation of a Merkle tree for localized integrity checks. The implementation of OLAP databases, in general, simplifies the process of generating encrypted backups due to the reduced likelihood of data modifications. However, while analytical workloads often operate on static data, efficiently handling updates is still an open research area. 

Sensitive information is thus encrypted both at rest on disk and in memory, with decryption occurring exclusively within CPU caches, emitting decrypted (aggregated) results. Our related work on Responsible Decentralized Data Architectures (RDDA)~\cite{battiston2023improving} elaborates on how we design our system to guarantee user privacy at the data definition and manipulation levels, leveraging SGX as a secure computation technology to emit anonymized query results.

However, additional measures are required to provide a fully secure system, particularly in safeguarding data in use. These may include hiding access patterns, encrypting buffers, preventing malicious signal injection, and implementing secure out-of-core execution. Running the application with hyperthreading disabled can mitigate certain well-known attacks\footnote{\href{https://sgx.fail/}{https://sgx.fail/}}, and the newer Intel hardware addresses most vulnerabilities, but as we previously mentioned, all these security measurements need to be promptly integrated. Moreover, all Parquet-related metadata must be protected to prevent the inference of their structure and row groups. We plan to address these enhancements in our future work.

Furthermore, database administrators will need to ensure adherence to security policies such as strict access control, authentication mechanisms, and user permissions. We acknowledge the presence of limited security controls and access policies in our database system. However, using encryption within DuckDB eases the management of access privileges by regulating access to encryption keys. Furthermore, the memory encryption in SGX mitigates the risks associated with potential memory leaks or buffer overflows. 

\section{Conclusion and Future Work}
We evaluated the confidential execution of analytical queries on DuckDB using SGX2 and Modular Parquet Encryption.
Our results show that a well-configured system only incurs a 1.5x-2x overhead, primarily due to the higher CPU cache miss cost in the enclave. 
However, there are performance hazards, mostly EPC paging, and a lack of NUMA locality.
We tuned DuckDB to use AES CPU instructions, significantly speeding up decryption.
Also, we disabled {\small\tt jemalloc}, which is SGX-unaware and causes EPC paging after memory fragmentation. 

There is still significant future work to be done, both in evaluating this setup on different enclave technologies and improving database architectures for TEEs. Previous work~\cite{lutsch2024benchmarking} has managed to gain performance improvements for query execution operations tailored specifically to SGX. However, this is only a small part of all the research opportunities opened by confidential computing within data management systems.

Furthermore, our prototype does not address some necessary security measurements such as the ones described in Section \ref{sec}, and its performance is not optimized for data transfer: data is encrypted twice when moved inside the enclave, which incurs significant overhead. Our roadmap includes implementing new system calls such that the decryption of the data happens only when necessary.

We advocate and plan for vectorized decryption, which requires more fine-grained encryption units than Parquet allows. We aim to only decrypt into CPU caches, thus avoiding EEM (Enclave Memory Manager) and memory movement costs. Finally, database systems could not only use enclave memory but also some unsafe memory to store non-sensitive or file-encrypted data, avoiding EEM overhead. Such improvements would also allow more portability and support of other secure hardware technologies while further improving the query performance in analytical workloads.

\begin{acks}
The authors would like to thank Intel Corporation, particularly Benny Fuhry and Dmitrii Kuvaiskii, for providing us with access to its hardware and technical support, and Adrian Lutsch for the feedback on this paper. 
\end{acks}

\newpage

\bibliographystyle{ACM-Reference-Format}
{\scriptsize
\bibliography{sample-base}}


\end{document}